\documentclass{aa}
\usepackage{graphics,epsf,epsfig,graphicx,natbib}

\bibpunct{(}{)}{;}{a}{}{,}

\begin{document}
\newcommand{\MHI}{\mbox{$M_{\mbox{\tiny HI}}$}}
\newcommand{\NHI}{\mbox{$N_{\mbox{\tiny HI}}$}}
\newcommand{\NHH}{\mbox{$N_{\mbox{\tiny H2}}$}}
\newcommand{\VHI}{\mbox{$V_{\mbox{\tiny HI}}$}}
\newcommand{\Vopt}{\mbox{$V_{\mbox{\tiny opt}}$}}
\newcommand{\NH}{\mbox{$N_{\mbox{\tiny H}}$}}
\newcommand{\Lb}{\mbox{$L_{\mbox{\tiny B}}$}}
\newcommand{\LbA}{\mbox{$L_{\mbox{\tiny B}}^{\mbox{\tiny A}}$}}
\newcommand{\LbB}{\mbox{$L_{\mbox{\tiny B}}^{\mbox{\tiny B}}$}}
\newcommand{\Lv}{\mbox{$L_{\mbox{\tiny V}}$}}
\newcommand{\Lr}{\mbox{$L_{\mbox{\tiny R}}$}}
\newcommand{\LHa}{\mbox{$L_{\mbox{\tiny H}\alpha}$}}
\newcommand{\Ha}{\mbox{H${\alpha}$}}
\newcommand{\Hb}{\mbox{H${\beta}$}}
\newcommand{\Hg}{\mbox{H$_{\gamma}$}}
\newcommand{\Hd}{\mbox{H$_{\delta}$}}
\newcommand{\Mb}{\mbox{$M_{\mbox{\tiny B}}$}}
\newcommand{\Mv}{\mbox{$M_{\mbox{\tiny V}}$}}
\newcommand{\Mr}{\mbox{$M_{\mbox{\tiny R}}$}}
\newcommand{\MHH}{\mbox{$M_{\mbox{\tiny H$_2$}}$}}
\newcommand{\HH}{\mbox{H$_2$}}
\newcommand{\Mo}{\mbox{M$_{\odot}$}}
\newcommand{\Mvir}{\mbox{$M_{\mbox{\tiny vir}}$}}
\newcommand{\Mdyn}{\mbox{$M_{\mbox{\tiny dyn}}$}}
\newcommand{\Mvis}{\mbox{$M_{\mbox{\tiny vis}}$}}
\newcommand{\Mmar}{\mbox{$M_{\mbox{\tiny mar}}$}}
\newcommand{\Lo}{\mbox{L$_{\odot}$}}
\newcommand{\Zo}{\mbox{Z$_{\odot}$}}
\newcommand{\x}{\mbox{$\times$}}
\newcommand{\Dph}{\mbox{$D_{25}$}}
\newcommand{\Vrot}{\mbox{$V_{\mbox{\tiny rot}}$}}
\newcommand{\rH}{\mbox{$r_{\mbox{\tiny H}}$}}
\newcommand{\Mk}{\mbox{$M_{\mbox{\tiny K}}$}}
\newcommand{\COa}{\mbox{CO(1$\rightarrow$0)}}
\newcommand{\COb}{\mbox{CO(1$\rightarrow$1)}}
\newcommand{\CO}{\mbox{CO}}
\newcommand{\ICO}{\mbox{I$_{\mbox{\tiny CO}}$}}
\newcommand{\cmm}{\mbox{cm$^{-2}$}}
\newcommand{\cmmm}{\mbox{cm$^{-3}$}}
\newcommand{\Mpc}{\mbox{Mpc}}
\newcommand{\kms}{\mbox{km~s$^{-1}$}}
\newcommand{\pcc}{\mbox{pc$^{-2}$}}
\newcommand{\kpc}{\mbox{kpc}}
\newcommand{\sbr}{\mbox{mag/\fbox{}\arcsec}}
\newcommand{\sbB}{\mbox{$\mu _{\mbox{\tiny B}}$}}
\newcommand{\sbV}{\mbox{$\mu _{\mbox{\tiny V}}$}}
\newcommand{\sbc}{\mbox{$\mu _{\mbox{\tiny B}0}$}}
\newcommand{\fB}{\mbox{$f_{\mbox{\tiny B}}$}}
\newcommand{\aop}{\mbox{$a_{25}$}}
\newcommand{\bop}{\mbox{$b_{25}$}}
\newcommand{\tabsp}{\noalign{\smallskip}}
\newcommand{\BV}{\mbox{$\mbox{B} - \mbox{V}$}}
\newcommand{\VK}{\mbox{$\mbox{V} - \mbox{K}$}}
\newcommand{\sigint}{\mbox{$\sigma _{\mbox{\tiny int}}$}}
\newcommand{\Hoa}{\mbox{H$_0= 75$~km~s$^{-1}$~Mpc$^{-1}$}}
\newcommand{\Hob}{\mbox{H$_0= 70$~km~s$^{-1}$~Mpc$^{-1}$}}
\newcommand{\qo}{\mbox{q$_0$}}
\newcommand{\Mbo}{\mbox{$M_{\odot}^{\mbox{\tiny B}}$}}
\newcommand{\Vpar}{\mbox{$V_{\mbox{\tiny par}}$}}
\newcommand{\eqw}{\mbox{$W$(H$_{\beta}$)}}
\newcommand{\Ab}{\mbox{$A_{\mbox{\tiny B}}$}}
\newcommand{\Av}{\mbox{A$_{\mbox{\tiny V}}$}}
\newcommand{\noteb}{\mbox{$^{\mbox{\tiny +}}$}}
\newcommand{\mJyb}{\mbox{mJy~beam$^{-1}$}}
\newcommand{\mJy}{\mbox{mJy}}
\newcommand{\K}{\mbox{K}}
\newcommand{\Ts}{\mbox{$T_{\mbox{s}}$}}
\newcommand{\TOIII}{\mbox{$T_e({\mbox{OIII}}$)}}
\newcommand{\TOII}{\mbox{$T_e({\mbox{OIII}}$)}}
\newcommand{\micron}{\mbox{$\mu$m}}
\newcommand{\Oabun}{\mbox{$12 + \log(\frac{\mbox{O}}{\mbox{H}})$}}
\newcommand{\er}{\mbox{$\pm$}}

\newcommand{\fb}{\mbox{$f_{\mbox{\tiny B}}$}}
\newcommand{\Lfir}{\mbox{$L_{\mbox{\tiny FIR}}$}}
\newcommand{\Lir}{\mbox{$L_{\mbox{\tiny IR}}$}}
\newcommand{\uflux}{\mbox{erg~cm$^{-2}$~s$^{-1}$}}
\newcommand{\usbflux}{\mbox{erg~s$^{-1}$~cm$^{-2}$~arcsec$^{-2}$}}
\newcommand{\ufluxm}{\mbox{erg~cm$^{-2}$~s$^{-1}$~\AA$^{-1}$}}
\newcommand{\ul}{\mbox{erg~s$^{-1}$}}
\newcommand{\uSFR}{\mbox{M$_{\odot}$~yr$^{-1}$}}
\newcommand{\muJy}{\mbox{$\mu$Jy}}
\newcommand{\OIIIa}{\mbox{[OIII]$_{\lambda 4959}$}}
\newcommand{\OIIIb}{\mbox{[OIII]$_{\lambda 5007}$}}
\newcommand{\OIIIc}{\mbox{[OIII]$_{\lambda 4363}$}}
\newcommand{\OI}{\mbox{[OI]$_{\lambda 6300}$}}
\newcommand{\OII}{\mbox{[OII]$_{\lambda 3727}$}}
\newcommand{\OIIb}{\mbox{[OII]$_{\lambda 7320,30}$}}

\newcommand{\OIIIt}{\mbox{[OIII]}}
\newcommand{\OIIt}{\mbox{[OII]}}
\newcommand{\OIt}{\mbox{[OI]}}
\newcommand{\SIIt}{\mbox{[SII]}}
\newcommand{\NIIt}{\mbox{[NII]}}
\newcommand{\NIt}{\mbox{[NI]}}
\newcommand{\ArIIIt}{\mbox{[ArIII]}}

\newcommand{\NIIa}{\mbox{[NII]$_{\lambda 6548}$}}
\newcommand{\NIIb}{\mbox{[NII]$_{\lambda 6584}$}}
\newcommand{\SIIa}{\mbox{[SII]$_{\lambda 6717}$}}
\newcommand{\SIIb}{\mbox{[SII]$_{\lambda 6731}$}}
\newcommand{\SII}{\mbox{[SII]$_{\lambda 6717,6731}$}}

\newcommand\cola {\null}
\newcommand\colb {&}
\newcommand\colc {&}
\newcommand\cold {&}
\newcommand\cole {&}
\newcommand\colf {&}
\newcommand\colg {&}
\newcommand\colh {&}
\newcommand\coli {&}
\newcommand\colj {&}
\newcommand\colk {&}
\newcommand\coll {&}
\newcommand\colm {&}
\newcommand\coln {&}
\newcommand\eol{\\}
\newcommand\extline{&&&&&&&&&\eol}

%
\def\HI{H\,{\small I}}
\def\HI{H\,{\sc i}}
\def\HII{H\,{\sc ii}}
\def\HIit{\mbox{H\hspace{0.155 em}{\footnotesize \it I}}}
\def\nan{Nan\c{c}ay}
\def\sbu{${\rm mag\,\,arcsec^{-2 }} $ \ }
\newcommand{\am}[2]{$#1'\,\hspace{-1.7mm}.\hspace{.0mm}#2$}
\newcommand{\as}[2]{$#1''\,\hspace{-1.7mm}.\hspace{.0mm}#2$}
\newcommand\btab[5]{\begin{table*}[#1]{\parbox{#4}{\caption{#2}}\rule[-0.5ex]{0cm}{0.5ex} }
\begin{tabular*}{#4}{#5} }
\newcommand\sbtab[5]{\begin{table}[#1]{\parbox{#4}{\caption{#2}}\rule[-0.5ex]{0cm}{0.5ex} }
\begin{footnotesize}
\begin{tabular*}{#4}{#5} }
\newcommand{\etab}[4]{
\end{tabular*}
\vspace*{#1}
\begin{flushleft}
\parbox{#2}{#3}
\end{flushleft}
\label{#4}
\end{table*} }
\def\kato{\rule[-1.25ex]{0cm}{1.25ex}}
\def\pano{\rule[0.0ex]{0cm}{2.5ex}}
%
%
%
%
%
%


\def\rf@jnl#1{{#1}}
\def\aj{\rf@jnl{AJ }}                   
\def\araa{\rf@jnl{ARA\&A }}             
\def\apj{\rf@jnl{ApJ }}                 
\def\apjl{\rf@jnl{ApJ }}                
\def\apjs{\rf@jnl{ApJS }}               
\def\ao{\rf@jnl{Appl.~Opt.}}           
\def\apss{\rf@jnl{Ap\&SS }}             
\def\aap{\rf@jnl{A\&A }}                
\def\aapr{\rf@jnl{A\&A~Rev.}}          
\def\aaps{\rf@jnl{A\&AS }}              
\def\azh{\rf@jnl{AZh }}                 
\def\baas{\rf@jnl{BAAS }}               
\def\jrasc{\rf@jnl{JRASC }}             
\def\memras{\rf@jnl{MmRAS }}            
\def\mnras{\rf@jnl{MNRAS }}             
\def\pra{\rf@jnl{Phys.~Rev.~A}}        
\def\prb{\rf@jnl{Phys.~Rev.~B}}        
\def\prc{\rf@jnl{Phys.~Rev.~C}}        
\def\prd{\rf@jnl{Phys.~Rev.~D}}        
\def\pre{\rf@jnl{Phys.~Rev.~E}}        
\def\prl{\rf@jnl{Phys.~Rev.~Lett.}}    
\def\pasp{\rf@jnl{PASP }}               
\def\pasj{\rf@jnl{PASJ }}               
\def\qjras{\rf@jnl{QJRAS }}             
\def\skytel{\rf@jnl{S\&T }}             
\def\solphys{\rf@jnl{Sol.~Phys.}}      
\def\sovast{\rf@jnl{Soviet~Ast.}}      
\def\ssr{\rf@jnl{Space~Sci.~Rev.}}     
\def\zap{\rf@jnl{ZAp }}                 
\def\nat{\rf@jnl{Nature }}              
\def\iaucirc{\rf@jnl{IAU~Circ.}}       
\def\aplett{\rf@jnl{Astrophys.~Lett.}} 
\def\apspr{\rf@jnl{Astrophys.~Space~Phys.~Res.}}
\def\bain{\rf@jnl{Bull.~Astron.~Inst.~Netherlands}} 
\def\fcp{\rf@jnl{Fund.~Cosmic~Phys.}}  
\def\gca{\rf@jnl{Geochim.~Cosmochim.~Acta}}   
\def\grl{\rf@jnl{Geophys.~Res.~Lett.}} 
\def\jcp{\rf@jnl{J.~Chem.~Phys.}}      
\def\jgr{\rf@jnl{J.~Geophys.~Res.}}    
\def\jqsrt{\rf@jnl{J.~Quant.~Spec.~Radiat.~Transf.}}
\def\memsai{\rf@jnl{Mem.~Soc.~Astron.~Italiana}}
\def\nphysa{\rf@jnl{Nucl.~Phys.~A}}   
\def\physrep{\rf@jnl{Phys.~Rep.}}   
\def\physscr{\rf@jnl{Phys.~Scr}}   
\def\planss{\rf@jnl{Planet.~Space~Sci.}}   
\def\procspie{\rf@jnl{Proc.~SPIE}}   

\let\astap=\aap
\let\apjlett=\apjl
\let\apjsupp=\apjs
\let\applopt=\ao

\title{VCC~2062: an old Tidal Dwarf Galaxy in the Virgo Cluster?}

\author{Pierre-Alain Duc\inst{1} \and Jonathan Braine\inst{2} \and Ute Lisenfeld\inst{3} \and Elias Brinks\inst{4} \and M\'ed\'eric Boquien\inst{1}}

\offprints{Pierre-Alain Duc, \email{paduc@cea.fr}}

\institute{Laboratoire AIM, CEA/DSM - CNRS - Universit\'e Paris Diderot, 
DAPNIA/Service d'Astrophysique, CEA-Saclay, F-91191 Gif-sur-Yvette Cedex, France
\and Observatoire de Bordeaux, UMR 5804, CNRS/INSU, B.P. 89, F-33270 Floirac, France
\and Dept. de F\'\i sica Te\'orica y del Cosmos, Universidad de Granada, Granada, Spain
\and Centre for Astrophysics Research, University of Hertfordshire, College Lane, Hatfield AL10 9AB, UK}

\date{Accepted: 31/08/2007}

\abstract
{Numerical simulations predict the existence of old Tidal Dwarf Galaxies (TDGs) that would have survived several Gyr after the collision lying at their origin. Such survivors, which would by now have become independent relaxed galaxies, would be ideal laboratories, if nearby enough, to tackle a number of topical issues, including the distribution of Dark Matter in and around galaxies. However finding old dwarf galaxies with a confirmed tidal origin is an observational challenge.}
{A dwarf galaxy in the nearby Virgo Cluster, VCC~2062, exhibits several unusual properties that are typical of a galaxy made out of recycled material. We discuss whether it may indeed be a TDG.}
{We analysed multi-wavelength observations of VCC~2062, including a CO map acquired with the IRAM 30m dish, an optical spectrum of its HII regions, GALEX ultraviolet and archival broad-band and narrow-band optical images as well as a VLA HI datacube, originally obtained as part of the VIVA project.}
{VCC~2062 appears to be the optical, low surface brightness counterpart of a kinematically detached, rotating condensation that formed within an HI tail apparently physically linked to the disturbed galaxy NGC~4694. In contrast to its faint optical luminosity, VCC~2062 is characterised by strong CO emission and a high oxygen abundance more typical of spiral disks. Its dynamical mass however, is that of a dwarf galaxy.}
{VCC~2062 was most likely formed within a pre-enriched gaseous structure expelled from a larger galaxy as a result of a tidal interaction. The natural provider for the gaseous tail is NGC~4694 or rather a former companion which subsequently has been accreted by the massive galaxy. According to that scenario, VCC~2062 has been formed by a past tidal encounter. Since its parent galaxies have most probably already totally merged, it qualifies as an old Tidal Dwarf Galaxy.}

\authorrunning{Duc et al.}
\titlerunning{An old TDG in the Virgo Cluster}
\maketitle

\section{Introduction}
The origin and properties of the satellites surrounding massive galaxies has recently been the subject of an active debate, triggered by its cosmological implications, new constraints from numerical simulations \citep[e.g.][]{Mayer07} as well as prolific observations.
In particular, optical surveys with cameras offering large fields of view allowed a deeper census of the faint dwarf satellite population around the Milky Way and Andromeda to be made \citep[e.g.][]{Belokurov07}. Although the number of known satellites keeps increasing with time, it is still lower than the number of primordial satellites predicted by standard cosmological hierarchical scenarios.  The situation could even be worse, as claimed by \cite{Bournaud06}. These authors have pointed out that the cosmological models do not take into account the fact that second-generation (or recycled) galaxies may be formed during collisions. According to the numerical simulations by \cite{Bournaud06}, a fraction of the so-called Tidal Dwarf Galaxies (TDGs), made out of tidal material expelled from parent colliding galaxies, survives long enough to contribute significantly to the population of satellites around massive hosts \citep[see also][]{Metz07}.

The formation in tidal tails of massive gravitationally bound objects has been known now for over a decade from observations of interacting systems \citep[see ][ and references therein]{Duc07}. However, so far only young TDGs have been unambiguously identified thanks to tidal arms and bridges linking them to their parent galaxies. As these features disappear in the course of time (typically of order 0.5--1 Gyr), galaxies of tidal origin become more difficult to distinguish from classical ones. Observing an old TDG existing as an independent entity several hundred Myr after the beginning of the collision lying at its origin, is a real challenge. This is because two defining characteristics --- a relative high metallicity inherited from their parent galaxies, and the absence of a prominent dark matter halo surrounding them \citep{Hunter00} --- can only be determined for nearby objects. Indeed, a good sensitivity is required to get the spectrophotometric data that allows a precise measurement of the oxygen abundances, while a high spatial resolution is indispensable to measure their dynamical mass and thus their dark matter content.
 The first TDG candidates so far studied in detail are the ones in the Antennae system (\object{NGC~4038/39}) at a distance of about 20 Mpc \citep{Mirabel92}. However doubts were raised whether they are self-gravitating objects \citep{Hibbard01}.
The nearby M81/M82/NGC~3077 group, at a distance of about 4 Mpc, shows instances of extra-planar, inter-galactic star-forming regions along or close to the HI tidal tails that link the three main colliding galaxies \citep[e.g.][]{Walter06}. Some of them, such as the Garland object or Holmberg IX, may qualify as TDG candidates \citep{Makarova02} although their dynamical status and real age are still unknown. Finally, even more nearby, a tidal origin has been speculated for the old Local Group dwarf spheroidals by \cite{Kroupa97} and \cite{Metz07}; however whether they really have a relatively low dark matter content, as claimed by these authors, is highly controversial \citep{Mateo98}.

At a mean distance of 17 Mpc, the Virgo Cluster, although it is poor in tidally interacting systems\footnote{\object{NGC~4438/35} is the only clear case of a recent major collision exhibiting strong morphological disturbances; \object{NGC~4567/68} shows two overlapping disks but no evident tidal features.}, may actually provide examples of galaxies made out of recycled material.
The cloud of molecular gas lying external to the NGC~4438/35 system \citep{Combes88} feeds intergalactic star formation \citep{Boselli05}. Tiny, point-like, intracluster HII regions have been reported near \object{NGC~4402} \citep{Cortese04} and \object{NGC~4388} \citep{Gerhard02}. Finally, isolated HI clouds of $10^7 - 10^8 ~\Mo$ have been found near NGC~4388 \citep{Oosterloo05}, \object{NGC~4254} \citep{Minchin05,Minchin07} and at several other locations in Virgo \citep{Kent07}. They were probably expelled from nearby spirals either by ram-pressure or tidal interactions\footnote{Note that the interpretation of the structure near NGC~4254, VirgoHI21, as a ``dark'' pristine galaxy by \cite{Minchin05} has been challenged among others by \cite{Bekki05} and more recently by \cite{Haynes07} and \cite{Duc07a}} and could as such be the progenitors of recycled galaxies. However their present HI column density is too low to allow star formation.

All these TDG candidates in Virgo were too young to have yet formed a significant stellar population and thus be included in the list of cluster members \citep[the Virgo Cluster Catalog, by][]{Binggeli85b}. One object in the VCC list, however, has intriguing properties that could be consistent with it being made of recycled material: VCC~2062. Originally considered to be a dwarf elliptical, it was later re-classified as a dwarf irregular when it was found to contain HII regions \citep{Sabatini05}. Its large HI mass with respect to its blue luminosity \citep{Hoffman96,Conselice03} is unusual. It actually lies within an HI structure that extends out to a nearby galaxy, \object{NGC~4694} \citep{Cayatte90, vanDriel89}.

We present here additional data to constrain the origin of \object{VCC~2062}: in particular, the metallicity of its HII regions and its CO emission. The paper is organised as follows: in Section~\ref{Sec:obs}, we present the radio, millimetric and optical observations of the system. The main results are described in Section~\ref{Sec:res} where special emphasis is given to the large scale and local environment of the object. Finally, the various hypotheses for the origin of VCC~2062 are discussed in Section~\ref{Sec:disc}.

Throughout this paper, we assume a distance of 17 Mpc. This value corresponds to the distance of the two massive galaxies closest to VCC~2062 \citep{Mei07}, as recently determined using the method of surface brightness fluctuations. At this distance, 1 arcmin corresponds to 4.9 kpc.

\section{Observations, data reduction and analysis}
\label{Sec:obs}

We have collected multi-wavelength data on the system NGC~4694 + VCC~2062. They were acquired through PI programmes or retrieved from archival databases.

\subsection{HI observations}
The NRAO\footnote{The National Radio Astronomy Observatory is a facility of the National Science Foundation operated under cooperative agreement by Associated Universities, Inc.} Very Large Array (VLA) data used in this paper were retrieved from the NRAO archival database. They had originally been obtained as part of the VIVA project (project ID: AK563), a systematic survey of HI-rich objects in Virgo \citep{Chung05,Chung07}. The observations of the NGC\,4694/VCC\,2062 system were made on 13/14 May 2004. Almost 8\,hr were spent on target interrupted at 25\,min intervals by brief scans on the primary calibrator 3C286 (J1331+3030). Given the proximity on the sky of this calibrator with the target, it was used for both amplitude and phase calibration.

The VLA correlator was used in mode 2BD employing a bandwidth of 3.125\,MHz and 128 channels, resulting in a 5.2~\kms\ channel spacing. As no on-line Hanning smoothing was applied, neighbouring channels are not fully independent and the effective resolution is 7.3~\kms. The centre of the band was set at a redshift corresponding to 1117~\kms.

We used the AIPS data reduction package and standard procedure to calibrate the data and produce a continuum emission subtracted spectral line data cube. We made cubes using different weightings by varying the {\sc robust} parameter in the AIPS task {\sc imagr}. Given the faintness of the emission we opted for maps with the highest sensitivity (equivalent to natural weighing). The beam size in these maps is almost circular at $18.^{\prime\prime}7 \times 17.^{\prime\prime}1$ (beam position angle: $-9^{\circ}$). The rms noise is 0.51\,mJy\,beam$^{-1}$. This corresponds to 0.97\,K and a detection threshold (assuming a $3\,\sigma$ detection across 3 channels) of $9 \times 10^{19}$\,cm$^{-2}$.

\begin{figure*}[!htbp]
\centering
\includegraphics[width=\textwidth]{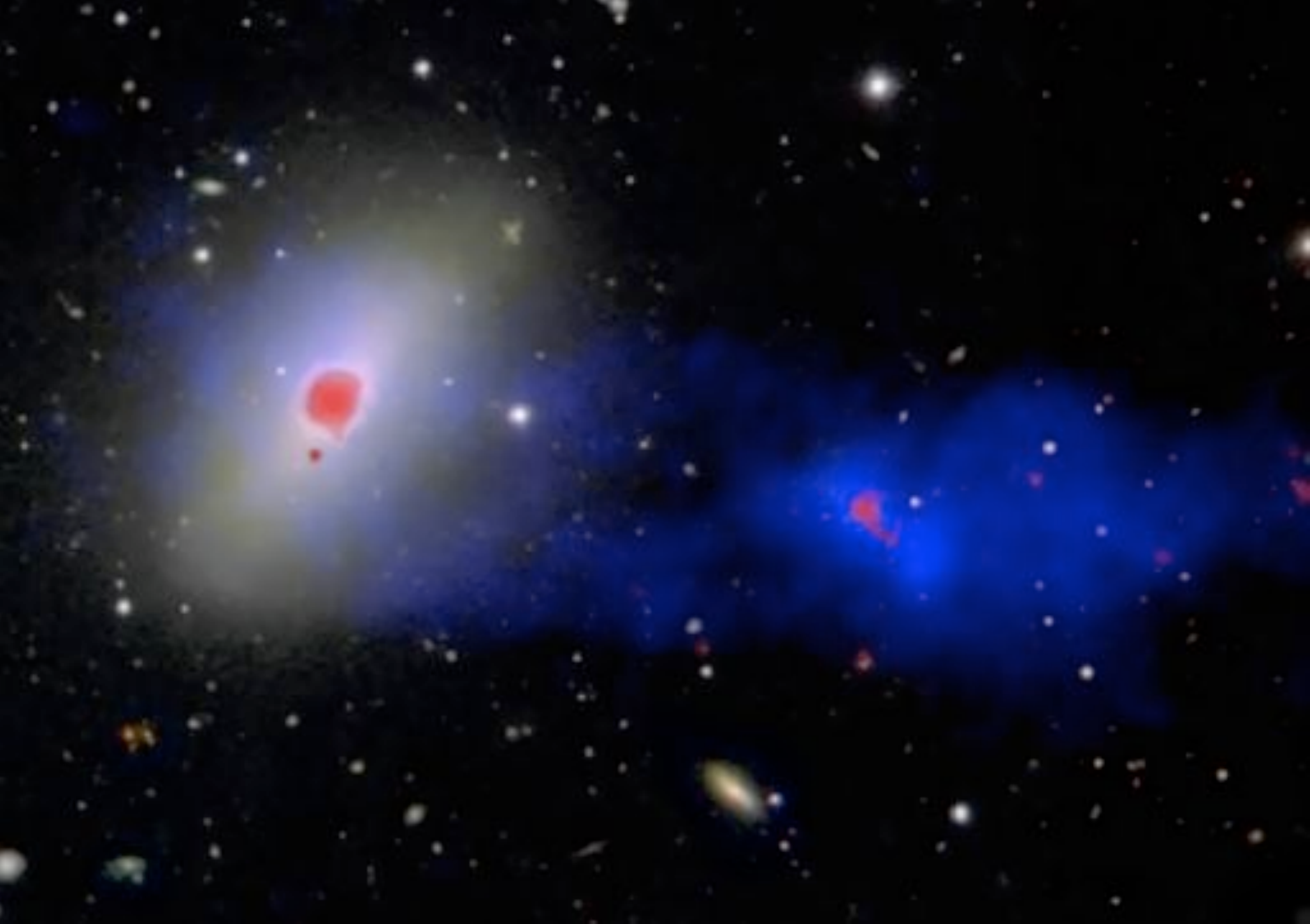}
\caption{VLA map of the HI gas distribution (in blue) around NGC~4694 (to the left) and VCC~2062 (to the right) superimposed on a true colour ($BVR$) optical image of the system. The GALEX-FUV emission, tracing regions of recent star formation, is overlaid in red. The field of view is $9^\prime \times 6^\prime$. North is up and East to the left. \label{fig:system}}
\end{figure*}

The distribution of the HI gas around NGC~4694 and VCC~2062 is shown in Fig.~\ref{fig:system}.  The HI map was produced as follows: we convolved  the natural--weighted data cube to $30^{\prime\prime}$ spatial resolution and blanked it at the $2\sigma$ noise level, only accepting those regions which contain emission in at least 3 consecutive channels. We then used this cube to apply conditional blanking to the original  robust and natural cubes and produced integrated maps (AIPS task {\sc xmom)}.

\subsection{CO observations}
\label{CO:obs}
\begin{figure*}[!htbp]
\centering
\includegraphics[width=\textwidth]{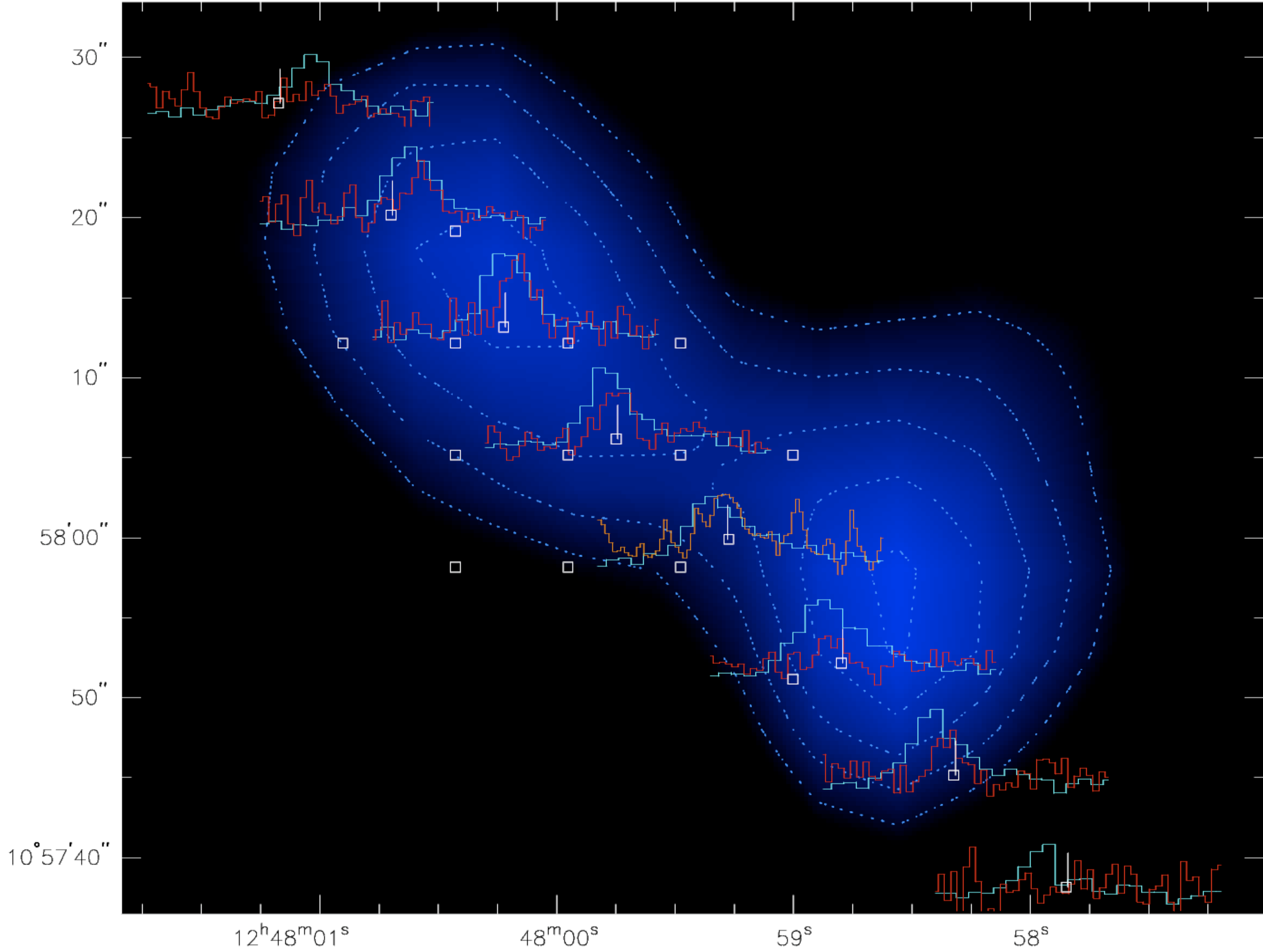}
\caption{A selection of IRAM CO(1--0) (red lines) and VLA HI (cyan lines) spectra along the main HI structure encompassing VCC~2062 and HI/SW. The small boxes indicate the positions of all IRAM pointings, including those not displayed here.  The filled box corresponds to the local HI peak. For one position, the CO(1--0) spectrum could not be acquired and the CO(2--1) line is shown instead in orange. The velocity range of all HI and CO spectra is 1090--1210 \kms. The vertical tick  indicates the systemic velocity of 1145 \kms.  The same vertical scale is used for all CO(1-0) spectra. The intensity of the line reaches a maximum of 0.06~K. The background image consists of the smoothed HI intensity map with its contour levels displayed in blue. The lowest contour corresponds to an HI column density of $7\x10^{20}~\cmm$. The CO(1-0) and HI observations lead to roughly similar beam sizes of about $20^{\prime\prime}$. The axes of the figure are in J2000.0 equatorial coordinates. \label{fig:VCC-COdisp-all}}
\end{figure*}

A CO map towards the densest regions of the HI tail has been compiled using data collected at Pico Veleta (Spain) with the 30 meter millimetre-wave telescope run by the Institut de Radio Astronomie Millim\'etrique (IRAM). The observations were carried out in August 2003 and November 2006 as part of a project to study systematically the molecular gas content of Tidal Dwarf Galaxy candidates \citep[][2001]{Braine00}\nocite{Braine01}. During the run in August 2003 we observed positions offset by $7^{\prime\prime}$ centered on the HI cloud associated with the optical galaxy VCC~2062. During the subsequent observing run in November 2006 we obtained a set of spectra (also spaced at $7^{\prime\prime}$), roughly aligned with the major axis of the HI structure, covering both the VCC~2062 cloud and another gaseous condensation to the South-West, named hereafter HI/SW.
The observed positions are indicated in Fig.~\ref{fig:VCC-COdisp-all} .
The CO(1-0) and CO(2-1) transitions at 115 and 230 GHz respectively were observed simultaneously and in both polarisations. A bandwidth of over 1300 \kms\ was available in both transitions using the two 512 \x 1 MHz filterbanks at 115 GHz and the two 256 \x 4 MHz filterbanks at 230 GHz.
System temperatures were typically 250 -- 300 K for the CO(1-0) transition and 300 -- 400 K for the CO(2-1) transition. The forward (main beam) efficiencies at Pico Veleta are currently estimated at 0.95 (0.74) at 115 GHz and 0.91 (0.54) at 230 GHz and the half-power beamwidths are about $21^{\prime\prime}$ and $11^{\prime\prime}$. All observations were done in wobbler switching mode, usually with a throw of $100^{\prime\prime}$ in azimuth, in order to be sure not to have emission in the reference beam.

The individual CO(1--0) spectra along the major HI axis are presented in Fig.~\ref{fig:VCC-COdisp-all}, together with the HI spectra at the same positions. For clarity, the off-major axis spectra are not shown; they have however been taken into account in the molecular gas mass calculation.

\subsection{Ultraviolet and optical imaging}
The UV satellite GALEX observed the field around NGC~4694 in May 2007. Images were obtained in the NUV (near-ultraviolet; $\lambda_{\rm eff}=227$ nm) and FUV (far-ultraviolet; $\lambda_{\rm eff}=151$ nm) filters. The field of view had a diameter of 1.24\degr and the PSF (point spread function) had a width of $\sim$5.0\arcsec.  The exposure times were 1530 seconds for each filter. The FUV image of the system is shown in Fig.~\ref{fig:system}.

\begin{figure}
\centering
\includegraphics[width=\columnwidth]{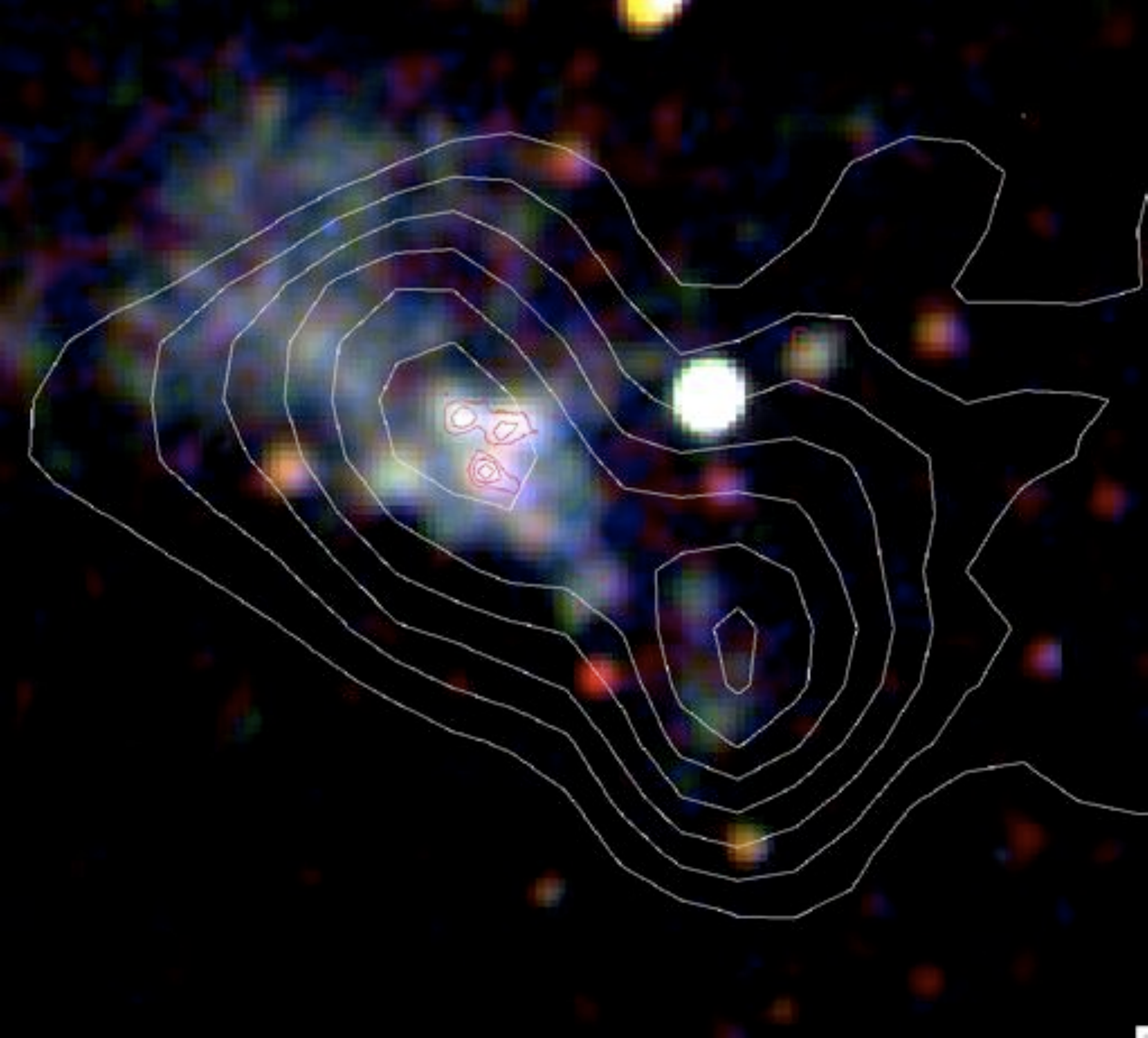}
\caption{True colour ($BVR$) optical image of VCC~2062 and HI/SW. The HI (white contours; lowest $4.0 \x 10^{20}$ cm$^{-2}$; intervals $1.0 \x 10^{20}$ cm$^{-2}$) and H$\alpha$ (red contours) distributions are superimposed. The field of view is $1.^{\prime}7 \times 1.^{\prime}5$.  North is up and East to the left.\label{fig:VCC-tcol}}
\end{figure}

 Optical, $BVR$, images of the field around VCC~2062 have been obtained in July 1994 with the EMMI instrument installed on the ESO NTT at La Silla Observatory. The pixel size was 0.27\arcsec, the average seeing 1.2\arcsec and the field of view, 8.5\arcmin $\times$ 9\arcmin. The exposure times were 300 seconds for each filter.
Images with a larger field of view, encompassing NGC~4694, were obtained from the Sloan database. A composite, true-colour image of VCC~2062 is shown in Fig.~\ref{fig:VCC-tcol}.

Aperture photometry was carried out on all these images using the IRAF {\em digiphot} package. Special care was taken in eliminating foreground stars or background galaxies identified on the basis of their compact size and colours as compared to the rest of the system \citep[see details of the method in][]{Boquien07}. As its surface brightness is particularly low, the background subtraction was challenging, resulting in large uncertainties in the measured fluxes, especially for the SDSS images.  As in \cite{Boquien07}, the background level was measured manually at several locations around the object and then averaged. The absolute flux calibration of the EMMI observations suffered from mixed weather conditions; we thus decided not to use the photometric standard stars observed during the run. Instead we derived the photometric zero point using several stars in the field with available SDSS fluxes transforming the Sloan system into the standard $BVR$ one (with the calibration of Lupton 2005, presented in the SDSS web pages). An independent photometric calibration has been carried out using a shallow image of the same field acquired in May 2007 with the 182cm Copernico telescope at Asiago Observatory.   The inferred value of the zero point differs by less than 0.05 mag for the V band and 0.15 mag for the B band.
The derived fluxes (not corrected for Galactic extinction) are listed in Table~\ref{tab:photometry}.
Note that the magnitudes of VCC~2062 published in the literature show a large dispersion, probably due to the different ways the foreground stars have (or haven't) been subtracted.

\begin{table}[htdp]
\caption{VCC~2062: broad and narrow band fluxes}
\begin{center}
\begin{tabular}{ccc}
\hline
\hline
Band & Central $\lambda$ & Flux\\
           &   ($\mu$m) & (mJy) \\
\hline
FUV (GALEX) & 0.151  & 0.04 $\pm$ 0.01 \\
NUV (GALEX) & 0.227 & 0.05 $\pm$ 0.01 \\
u' (SDSS)  & 0.355  & 0.13: $\pm$ 0.09 \\
B (NTT) & 0.440  & 0.20 $\pm$  0.03  (B=18.32 mag)\\
 g' (SDSS) & 0.468  & 0.22 $\pm$ 0.04 \\
V (NTT) &  0.550  & 0.23 $\pm$ 0.02  (V=18.00 mag)\\
r' (SDSS) & 0.616  & 0.24 $\pm$ 0.06 \\
R (NTT) &  0.640 & 0.25 $\pm$ 0.03 (R=17.71 mag)\\
i' (SDSS) & 0.748  & 0.23 $\pm$ 0.07 \\
z' (SDSS) & 0.893  & 0.26: $\pm$ 0.20 \\
H$\alpha$ &  0.656  & $4 \x 10^{-15}~\uflux$\\ 
\hline
\end{tabular}
\end{center}
\label{tab:photometry}
\end{table}%

\subsection{Optical spectrophotometry}
\label{sect:spec}
Fully reduced narrow-band H$\alpha$ images acquired in 1995 with the 0.9m telescope at the Kitt Peak National Observatory were downloaded through the NED database. Flux calibration was performed by \cite{Koopmann01}.

\begin{figure}[!htbp]
\centering
\includegraphics[height=\columnwidth,angle=-90]{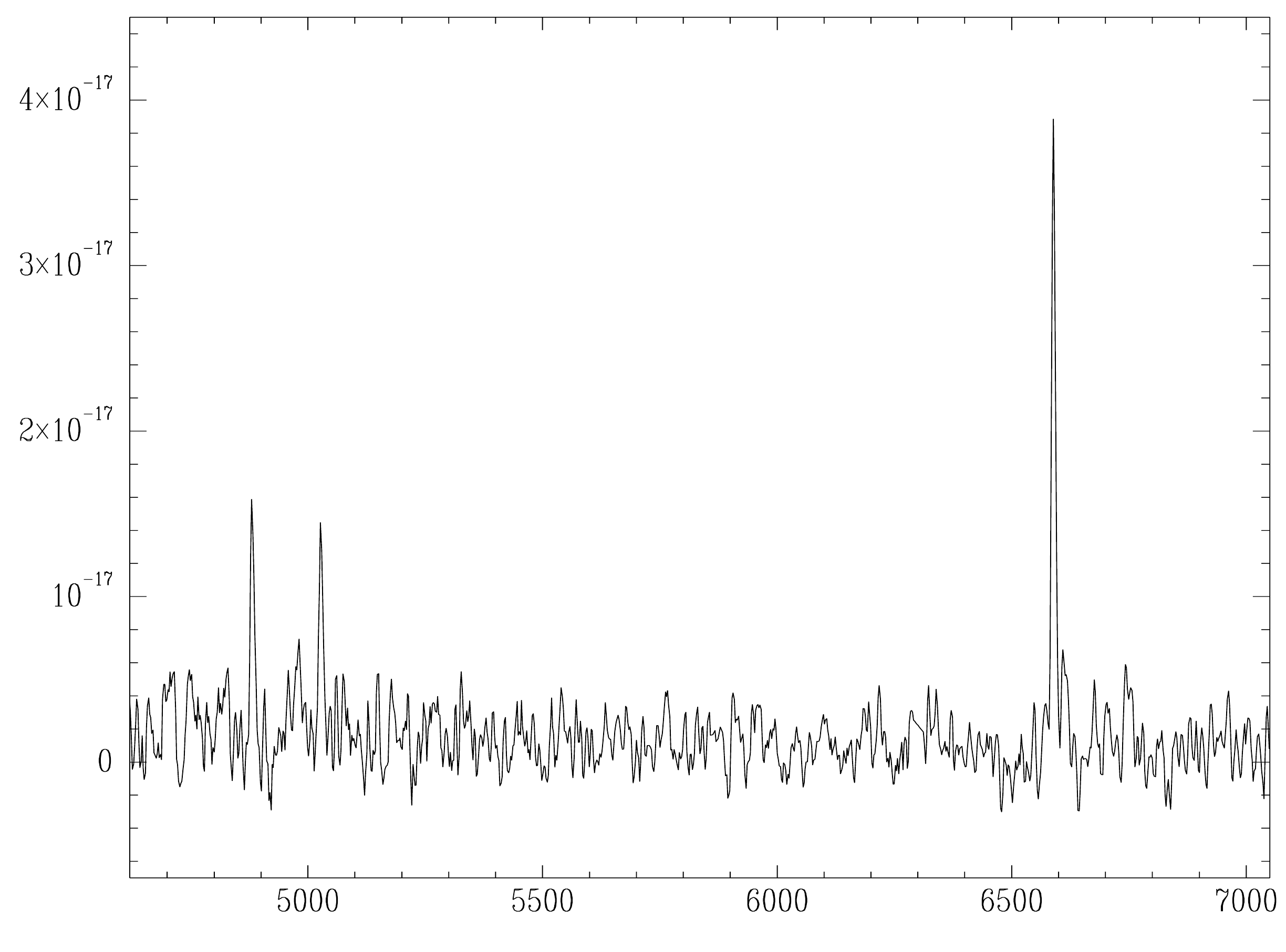}
\caption{Stacked optical spectrum of the HII regions of VCC~2062. The flux unit is in \ufluxm; the wavelength is in \AA. \label{fig:VCC-spec}}
\end{figure}

Multi-object spectroscopic observations were carried out in July 1994 with EMMI. Low resolution grism (ESO \#3) spectra were obtained towards 20 extended sources around NGC~4694 and VCC~2062. In addition, $1.^{\prime\prime}5$ wide slitlets were put on two of the three principle HII regions identified on the \Ha\ image of the object (see Fig.~\ref{fig:VCC-tcol}). To increase the signal to noise, these two spectra were averaged. The resulting spectrum is shown in Fig.~\ref{fig:VCC-spec}. The other slitlets were put on objects which turned out to be background galaxies and/or were too faint to exhibit any emission or absorption lines.

Line fluxes and errors were measured using the IRAF {\em splot} procedure. They are listed in Table~\ref{tab:spectrophotometry}.
The spectra were flux-calibrated using spectrophotometric standard stars observed throughout the run. The \Ha\ fluxes measured through the slits and grism and those derived from the narrow-band image were compared, correcting the former for aperture effects, and the latter for contamination by the \NIIa\ and \NIIb\ lines. They agree within a factor of 2, which is acceptable given the large uncertainty in the narrow-band calibration.
An estimate of the optical extinction of VCC~2062 was determined from the Balmer decrement, \Ha/\Hb. The inferred optical extinction being very low, less than 0.07 mag, no correction for dust extinction was applied to our (spectro-)photometric measurements.

\begin{table}[htdp]
\caption{VCC~2062: spectrophotometry}
\begin{center}
\begin{tabular}{c|c}
\hline
\hline
\Hb  & 100 $\pm$ 13  \\ 
\OIIIa &  60 $\pm$ 8 \\
\OIIIb & 93 $\pm$ 13 \\
\NIIa  & 26 $\pm$ 4 \\
\Ha & 291 $\pm$ 25 \\
\NIIb & 54 $\pm$ 7 \\
\SIIt & 71 $\pm$ 11 \\
\hline
\multicolumn{2}{p{5cm}}{\tiny\bf The line fluxes are normalized to the \Hb\ flux taken equal to 100 arbitrary units.}
\end{tabular}
\end{center}
\label{tab:spectrophotometry}
\end{table}%
\

Oxygen abundances were determined using two semi-empirical methods, as the lines required for using a more direct method were not detected. The classical R23 method could not be applied because the \OII\ line was outside our spectral range. We used instead the calibration of \cite{Edmunds84}, which relies on the sole measurement of the \OIIIa $+$ \OIIIb\ over \Hb\ flux ratio. This method is degenerate and we used the argument that the high \NIIb\ to \Ha\ flux ratio (see below) implies that the oxygen abundance should be read off from the upper branch of the metallicity versus oxygen line flux diagram. We obtained with this method $12+\log\left(\textrm{O/H}\right)=8.7$. Determining indirectly, but more reliably the oxygen abundance from the \NIIb\ to \Ha\ flux ratio \citep{Denicolo02}, we got a similar value of $12+\log\left(\textrm{O/H}\right)=8.6$. This is the value adopted in this paper; it is close to solar abundance\footnote{Taking $12+\log\left(\textrm{O/H}\right)=8.66$ for the solar metallicity \citep{Asplund05}}. Such a high metallicity is also suggested by the strong CO signal detected towards this object (see Sect.~\ref{Sect:gas}).

\section{Results}
\label{Sec:res}

We present here the environmental, global and local properties of VCC~2062, starting from the largest scales, and then zooming in to get a close-up of the object.

\subsection{Environment and global properties of the system NGC~4694 + VCC~2062}

\subsubsection{At the cluster scale}
\begin{figure*}
\centering
\includegraphics[width=\textwidth]{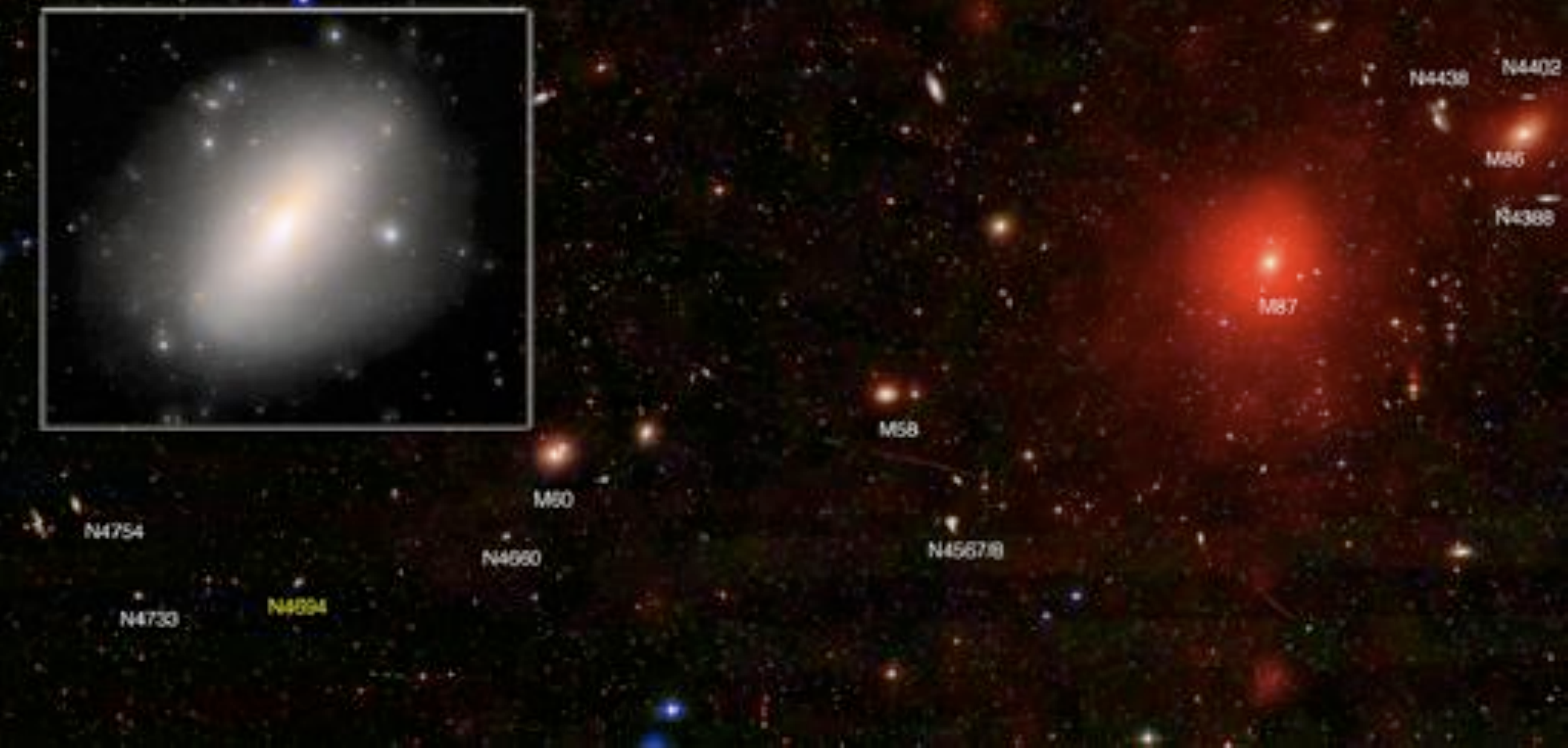}
\caption{The large scale environment of the system NGC~4694/VCC~2062. The principle galaxies and those mentioned in the text are labelled on this true colour SDSS image queried from the WIKISKY.com project.  The ROSAT all-sky X--ray map of the area is superimposed in red. The distance between NGC~4694 and M~87 is  about 4 degrees (1.1 Mpc). North is up and East to the left. The inset shows a close up of NGC~4694.  The low-surface brightness component of the halo has been enhanced.\label{fig:env}}
\end{figure*}

VCC~2062 lies in the outerskirts of the Virgo Cluster, at a distance of about 4 degrees (1.1 Mpc) to the East of the cluster core centered on M87, where the X-ray emission from the ICM peaks (see Fig.~\ref{fig:env}). This is a region with a rather low density in galaxies. The closest galaxy to the dwarf, $2.^\prime2$ South and slightly to the East, is a background object for which we measure a redshift of 0.06 (see Fig.~\ref{fig:system}).

The obvious environmental connection is with the perturbed galaxy NGC~4694 which lies at almost the same redshift (at 1175 \kms) and to which VCC~2062 is connected by a bridge of atomic hydrogen (see Fig.~\ref{fig:system}). NGC~4694 is classified in NED as an SB0 pec galaxy. The long HI feature stretching towards VCC~2062 is one of the many peculiarities of this early-type galaxy, along with the presence of dust lanes (see the inset of Fig.~\ref{fig:env}) and signs of recent star formation. The \HI\ bridge is about 38 kpc long (50 kpc when considering the faintest HI features to the West). VCC~2062 is located at about mid-distance. There are actually two parallel arms of low HI column density that connect the two galaxies, whereas further out to the West, the HI tail is more uniform and denser. The total mass of the HI structure,  as measured from the VLA beam corrected, natural weighing, map is $1.1 \x 10^9~\Mo$  including about $0.2  \x 10^9~\Mo$ associated with NGC~4694. The kinematics of the bridge is globally regular from outside NGC~4694 to the tip of the tail. In detail, some velocity substructures can be identified. They will be discussed later on.

Except for VCC~2062, the gaseous tail has no optical counterpart down to a brightness limit in the V band of $\sbV=26.3~\sbr$. However, the outer isophotes of NGC~4694 are asymmetric and show various protuberances  as seen in  Fig.~\ref{fig:env}.

\subsubsection{At the galaxy scale}
\label{sect:gal-scale}
We now zoom in, in the direction where the HI structure has its highest column densities, about 18 kpc West of NGC~4694. The tail there consists of two condensations with column densities above $5 \x 10^{20}$ cm$^{-2}$ and separated by 2.6 kpc (see Fig.~\ref{fig:VCC-tcol}). Only the slightly less massive one, to the North-East, has a clear optical counterpart: the object previously catalogued as VCC~2062. Very faint near-ultraviolet and optical emission with a surface brightness below $\sbV=25.5~\sbr$ can be seen towards the South-West HI clump, HI/SW.

The narrow-band \Ha\ image exhibits three distinct compact emission-line regions clustered at the HI peak of VCC~2062. Their integrated spectrum shown in Fig.~\ref{fig:VCC-spec} presents emission lines typical of star-forming (HII) regions. Far-ultraviolet emission is also observed at this location. However, outside these knots of star-formation, the stellar body has a very low surface brightness. The fitted central surface brightness is $\sbV=24.5~\sbr$. The morphology of VCC~2062 is very irregular, even showing a triangle shape at our sensitivity. No far-ultraviolet nor \Ha\ emission is detected towards HI/SW despite an HI column density above $10^{21}$ cm$^{-2}$ (see Fig.~\ref{fig:VCC-tcol}).

CO(1-0) and CO(2-1) molecular lines were detected along most of the densest regions of the HI structure, in particular towards HI/SW. They are however more prominent towards the star-forming regions of VCC~2062 (see Fig.~\ref{fig:VCC-COdisp-all}). The CO(1-0) to HI flux ratio varies by a factor of up to 4 between the HI peak of VCC~2062 and that of HI/SW.
The CO intensity is also found to decrease with distance from the major HI axis.

\begin{figure}[!htbp]
\centering
\includegraphics[width=\columnwidth]{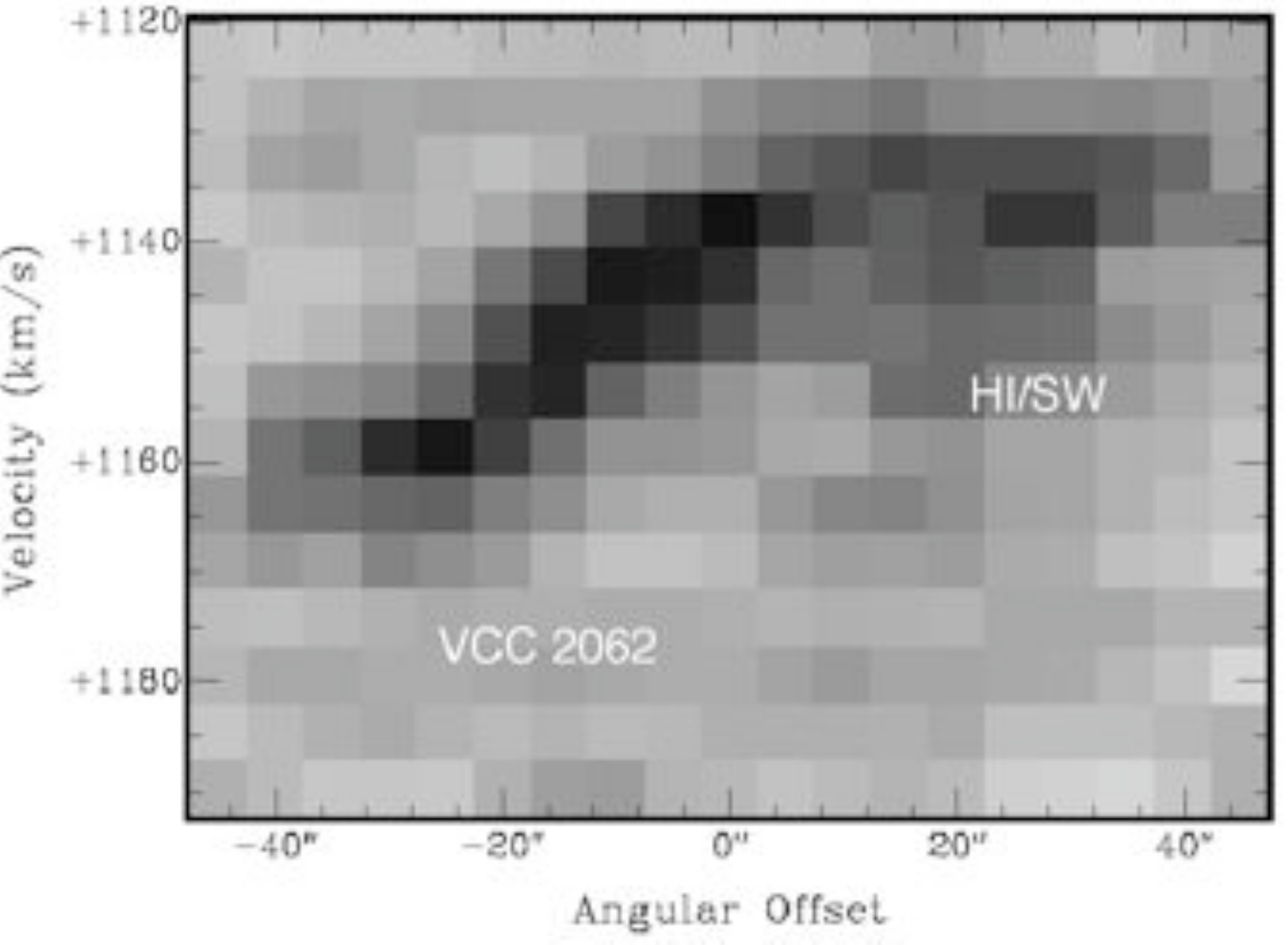}
\caption{HI Position-Velocity diagram along the major axis of VCC~2062, encompassing HI/SW (Position angle: 220 degrees).
 The PV diagram has been extracted from the HI datacube cube, putting the Position axis at an angle of 220 degrees, and spatially integrating the fluxes over a width of one pixel. The reference offset corresponds to the HI peak, as shown in Fig.~\ref{fig:VCC-COdisp-all}. Whereas VCC~2062 exhibits a clear velocity gradient, HI/SW shows broad \HI\ lines but, at our resolution, no spatial variation with velocity. \label{fig:PV}}
\end{figure}

Not only do VCC~2062 and HI/SW differ in their ability to form stars, but they also show different kinematics. A position-velocity diagram of the HI along the principle axis of VCC~2062 (as determined by its HI extent) shows a clear velocity gradient. When estimated from the HI envelope, it amounts to about 45~\kms, over a distance of 50 arcsec or about 4 kpc (see Fig.~\ref{fig:PV}). For comparison, the total velocity excursion along the entire HI tail is just 40~\kms. HI/SW does not show any velocity structure as a function of position. In the latter sub-structure, the HI line shows a broad and asymmetric profile, and a FWHM greater than 35~\kms, whereas it is 20~\kms\ towards VCC~2062. The kinematics derived from the CO(1-0) lines follows closely that derived from the HI (see Fig.~\ref{fig:VCC-COdisp-all})\footnote{Note that the velocity resolution of the IRAM CO spectra is about twice better than that of the VLA HI ones: 3~\kms\ instead of 7~\kms. However we do not have a full data cube to carry out a comprehensive kinematical study of the system in CO. Thus, in the following, we only consider the VLA HI cube for our kinematical analysis.}: when estimated from the peak value of the individual spectra (determined by fitting the lines with Gaussians), the velocity difference between the NE and SW side of VCC~2062 is exactly the same for the CO and HI: 15~\kms. However at a given position, a systematic offset between the  CO and HI of about 5 \kms\ is observed.  Besides the CO spectra have consistently smaller line widths, typically 15~\kms. Towards HI/SW, the signal to noise is too low to determine a reliable line width.

The fact that the HI is kinematically decoupled from the rest of the tail at the location of VCC~2062 suggests that the latter object may be dynamically independent. As a self-gravitating system, it qualifies as a galaxy, whatever its origin. In contrast, HI/SW has not (yet) reached a state of dynamical equilibrium, despite an HI mass and integrated column density slightly higher than VCC~2062.

\subsection{Physical properties of VCC~2062}

We discuss here the properties of VCC~2062 itself  and compare them with those of other 
(dwarf) galaxies in the Virgo Cluster.  Its main physical characteristics are summarised in Table~\ref{tab:prop}.  Note that all the integrated values presented in this section relate to the kinematical detached part of the HI structure, which as discussed before,  corresponds to what may be considered as an independent galaxy.

\begin{table}[htdp]
\caption{Integrated properties of VCC~2062}
\begin{center}
\begin{tabular}{c|c}
\hline
\hline
RA (J2000) & 12:48:00\\
DEC (J2000) & 10:58:14 \\
HI/CO heliocentric velocities & 1145 \kms \\
\Ha\ heliocentric velocity & 1180 \kms \\
Oxygen Abundance & 12+log(O/H)=8.6-8.7 \\
\Mb & -13.0 \\
HI Mass$^{*}$ & $7.0 \x 10^{7}~\Mo$ \\
H$_{2}$ Mass$^{*}$ & $1.7 \x 10^{7}~\Mo$   \\
SFR(H$\alpha$)$^{*}$ &  0.001 ~\uSFR\\
SFR(FUV)$^{*}$ &  0.002 ~\uSFR\\
\hline
\multicolumn{2}{p{7cm}}{\tiny\bf  $^*$ The values indicated here are integrated over the area corresponding to the kinematically detached part of the HI structure, which we refer as VCC~2062.}
\end{tabular}
\end{center}
\label{tab:prop}
\end{table}%

\subsubsection{Gas content}
\label{Sect:gas}
VCC~2062 contains about $7.0 \x 10^{7}~\Mo$ of atomic hydrogen, i.e. a bit less than 10\% of the total HI mass of the tail, not counting the component associated with NGC~4694.This measure takes into account all the mass associated with the kinematically decoupled component of the HI, and thus excludes HI/SW. Previous measurements of the HI mass of VCC~2062 \citep{Hoffman85,vanDriel89,Conselice03}, based on lower resolution observations, considered both components and were thus overestimates. 
The inferred HI mass to blue luminosity, \MHI/\Lb, of around 3~\Mo/\Lo\ is much lower than previously determined values, but still high with respect to the dwarf irregulars in the Virgo Cluster which have an average \MHI/\Lb\ of 0.6 \Mo/\Lo\ \citep{Conselice03}. It is in the range determined by \cite{Sabatini05} for a sample of Virgo Low Surface Brightness galaxies.

\begin{figure}[!htbp]
\centering
\includegraphics[width=\columnwidth,angle=0]{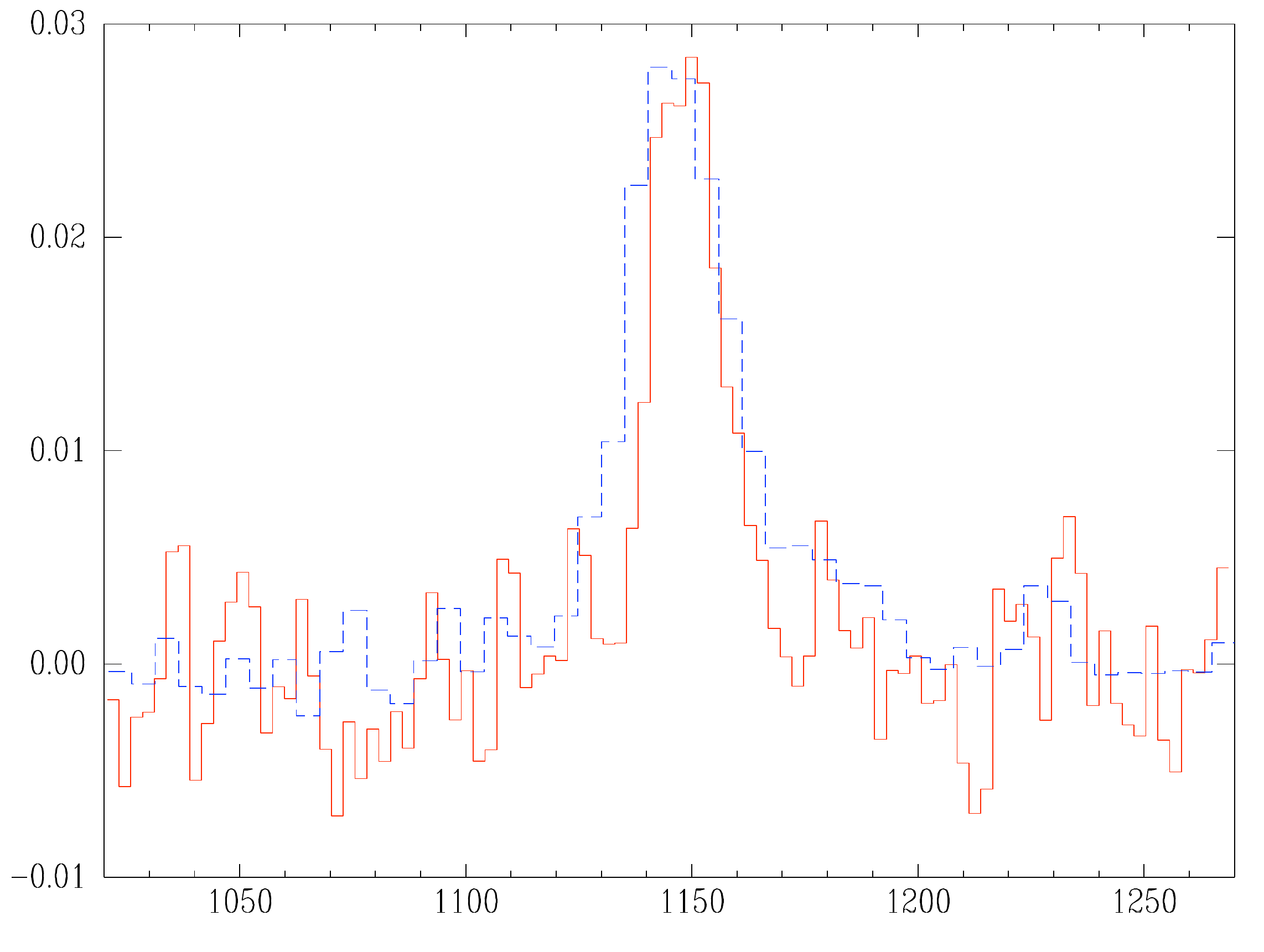}
\caption{IRAM CO(1-0) (red) and VLA HI (blue dashed) spectra of VCC~2062. The vertical scale gives the CO intensity in K and the horizontal one the velocity in \kms. The HI spectrum is in arbitrary units. The spectra have been averaged over a region corresponding to the optical body. \label{fig:VCC-COHIspec}}
\end{figure}

Compared to other Virgo galaxies of the same luminosity (\Mb=-13.0), VCC~2062 is actually most remarkable for its high molecular gas content, as derived from the intensity of the CO line. The CO emission towards VCC~2062  reaches an integrated intensity of 1 K \kms. This is the typical value measured for normal late-type galaxies in Virgo \citep{Boselli02} which are at least 50 times more massive. To our knowledge, VCC~2062 is the first low luminosity object in Virgo that has thus far been detected in CO. Assuming a standard galactic conversion factor of $\NHH/\ICO=2\x10^{20}$ \cmm\ \citep{Dickman86}\footnote{Since the metallicity of VCC~2062 is about solar, this conversion factor may also apply for this specific environment.}, we determined from the averaged measure of \ICO\ over the area of VCC~2062 (Fig.~\ref{fig:VCC-COHIspec}) an \HH\ mass of $1.7\x10^{7}~\Mo$ \citep[see details of the method to determine the mass in][]{Lisenfeld02}. The ratio of the \HH\ to HI mass of about 30\% is much higher than in classical dwarf galaxies and lies at the upper range of that of unperturbed massive galaxies in Virgo, i.e. those not affected by an HI deficiency. A typical value for Virgo spirals is 10\% \citep{Boselli02}.

\subsubsection{Stellar Populations}
\label{sect:stars}
The absolute blue magnitude of VCC~2062, $\Mb=-13.0$, clearly puts this object in the dwarf 
category, at least with respect to its overall stellar content. It was originally considered a dwarf elliptical, but its very irregular morphology on deep optical images clearly invalidates this classification. The fact that it contains active star-forming regions, traced by the compact HII regions detected towards it, is also at odds with it being a classical dE. The current Star Formation Rate (SFR), as derived from the \Ha\ luminosity and the conversion factor of \cite{Kennicutt98a}, is however low: about 0.001 ~\uSFR. The SFR derived from the GALEX FUV band, which is a measure for SF over a longer time scale of $\sim 100$\,Myr, reaches a consistent value of 0.002 \uSFR.
These estimates assume continuous star formation, with a Salpeter initial mass function and mass cut-offs below 0.1 and above 100 \Mo.
The gas consumption time or inverse of the star formation efficiency, defined as the ratio between the molecular gas mass and the Star Formation Rate, is as long as $\sim 10^{10}$ years, a factor of 5 higher than in local large spirals \citep{Kennicutt98a}.

\begin{figure}[!htbp]
\centering
\includegraphics[height=\columnwidth,angle=-90]{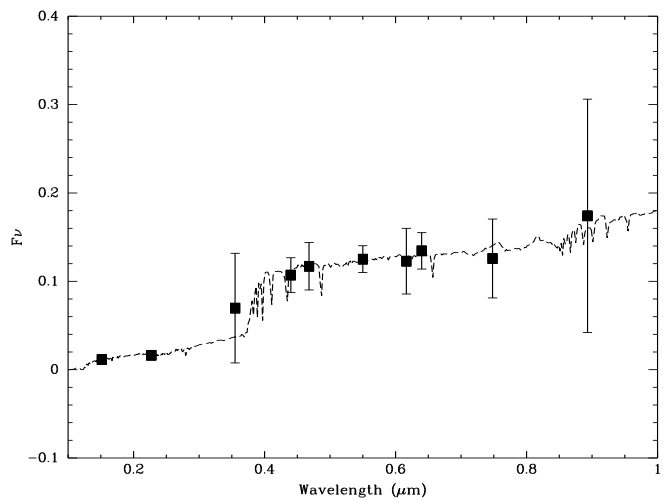}
\caption{Spectral Energy Distribution  of  a sub--region of VCC~2062 located to the North-East of its current star forming regions (see Fig.~\ref{fig:VCC-tcol}). Flux units are in mJy. Assuming a single burst model, the SED of the old stellar component is best fit by a model with an instantaneous burst which occurred about 0.3 Gyr ago. The corresponding synthesised spectrum is shown as a dashed curve. \label{fig:SED-regD}}
\end{figure}

The composite $BVR$ image of Fig.~\ref{fig:VCC-tcol} shows, except for the three knots of current star-formation, a rather uniform and blue colour of $B-V=0.35$. We used the evolutionary synthesis code PEGASE \citep{Fioc97} to roughly estimate the mean age of this background stellar population.  We carried out the aperture photometry of the North-Eastern region of VCC~2062 located outside its HII regions, in order not to be polluted by the latest star formation episodes. Our photometric database consists of 10 data points from the far-UV to the very near-infrared but lacks data in the near and mid-IR (see Fig.~\ref{fig:SED-regD}). With such a limited spectral range, the Spectral Energy Distribution can be fit with several scenarios for the Star Formation History. For reasons explained later, we tested the hypothesis that the bulk of the stars in the object (outside the current SF regions) were formed during a single episode of star formation. We varied the time of the onset of the starburst, the time scale of its decay and the extinction. We assumed a roughly solar metallicity, as suggested by the observations of the HII regions. The data are consistent with the hypothesis that the bulk of the background stars were formed about 0.3 Gyr ago and rule out a very old stellar component. The fitted extinction is low, which is consistent with that derived from the Balmer decrement in the HII regions. Of course more complex scenarios involving secondary bursts are still possible.

\subsubsection{Metallicity}
\begin{figure}[!htbp]
\centering
\includegraphics[width=\columnwidth]{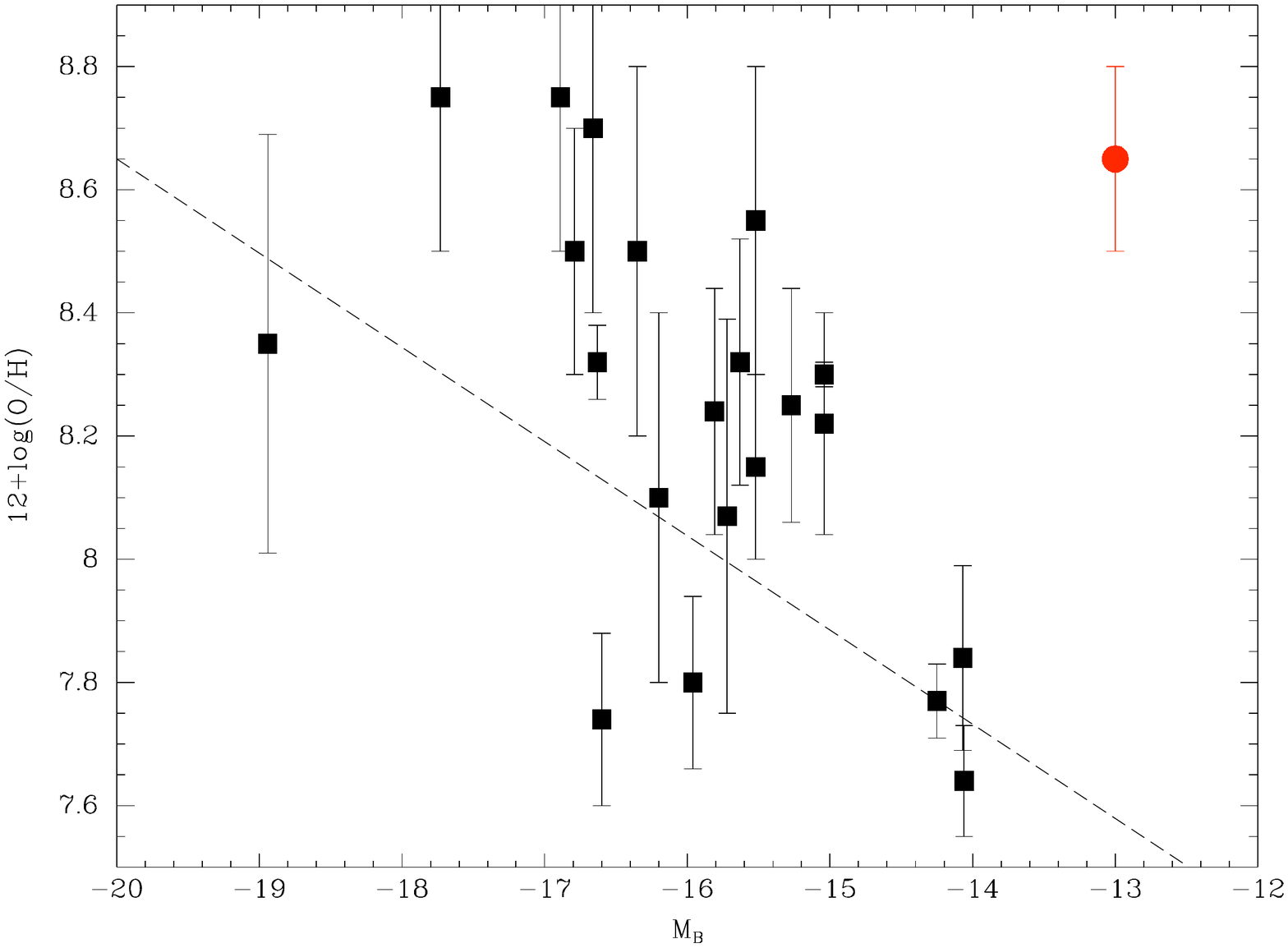}
\caption{Estimated metallicity versus absolute blue magnitude for a sample of dwarf galaxies in the Virgo Cluster. All oxygen abundances were taken from \cite{Lee03} and \cite{Vilchez03}. The error bars were taken from these references unless a range of metallicities was given. In that case, a systematic error of 0.1 dex has been added. The absolute blue magnitudes were taken from the Virgo GOLD Mine Database \citep{Gavazzi03}. Only objects at an estimated distance of 17 Mpc are included in this plot. VCC~2062 is shown as the red filled circle. The dashed line is the fit to the metallicity-luminosity relation established by \cite{Lee03} for a sample of nearby dwarf irregular galaxies. \label{fig:met-virgo}}
\end{figure}

As was argued earlier, the metallicity of VCC~2062 estimated from the emission lines of its HII regions, 12+log(O/H)=8.6-8.7, is roughly solar (see Sect.~\ref{sect:spec}). There is a well established relation between the oxygen abundance and the stellar mass, as approximated by the absolute blue magnitude. Given that relation, as calibrated for instance by \cite{Lee03} and the absolute magnitude of VCC~2062, one would have expected an oxygen abundance of 7.6, i.e. one dex below the measured value.
Oxygen abundances have been measured in a number of star-forming dwarf galaxies in the Virgo Cluster (see Fig~\ref{fig:met-virgo}). \cite{Vilchez95} noted a tendency for them to be over-abundant with respect to isolated dwarfs. This trend, clearly visible in Fig~\ref{fig:met-virgo}, has also been seen in other clusters \cite[e.g.][]{Duc01,Iglesias03} and can be accounted for by various hypotheses: confinement effects by the intra-cluster medium, disruption of previously more massive and more metal rich galaxies or a tidal origin.
Fig.~\ref{fig:met-virgo} indicates that, compared to the other dwarfs in Virgo,  VCC~2062 falls even more way off  the metallicity--luminosity relation.

\subsubsection{Dynamical and luminous masses}
Previous estimates of the dynamical mass of VCC~2062, as derived from the HI spectra, show an amazingly large range of values: from as large as $1.5 \x 10^{10}~\Mo$, and a ratio of $\Mdyn/ \Lb$ of about 300 \citep{Hoffman93}, to much more modest values of $\Mdyn=9 \x 10^{8}~\Mo$ and $\Mdyn/ \Lb$ of about 80 \citep{Conselice03}. Such discrepancies may partly be due to the fact that HI datacubes with different spatial and spectral resolution were used. The main source of uncertainty then comes from the way the HI was assumed to be associated with the dwarf. As indicated in Sect.~\ref{sect:gal-scale}, we only consider that HI component which appears kinematically detached from the tail, which seems to indicate to be in rotation, and which spatially coincides with the optical counterpart of VCC~2062. This component exhibits a well defined velocity gradient (see Fig.~\ref{fig:PV}) of 42 \kms\ over a diameter of 4.2 kpc. The velocities were determined from the envelope at 50\% of the HI peak level, as in \cite{Bournaud07}
and a ``background'' large-scale velocity gradient of 4 \kms\ across VCC~2062 was subtracted to correct for any streaming motions along the tail. We derive a range of dynamical masses of $\Mdyn=R\x (( \Vrot/sin(i))^2+ \sigma^2)/G =  3-4  \x 10^{8}~\Mo \pm 1  \x  10^{8}~\Mo$. We assumed a velocity dispersion $\sigma$ of the ISM of about 8~\kms\ (in agreement with the measured FWHM of $\sim 20$~\kms), and an inclination $i$ between 45 and 60 degrees, based on the optical morphology. The error takes into account the signal to noise of the HI and the systematic error of the method, as estimated from numerical simulations \citep[for details see][]{Bournaud07}.
This dynamical -- or total -- mass may be compared to the luminous one which is wholly dominated by the atomic hydrogen. Including a contribution by Helium, the neutral atomic component amounts to $1.0 \x 10^{8}~\Mo$  (see Sect.~\ref{Sect:gas}). The molecular mass inferred from the CO emission, assuming a galactic $\NHH/\ICO$ ratio, is $0.2 \x 10^{8}~\Mo$, again corrected for Helium. Simply assuming a stellar M/L ratio of 1, we infer from the blue luminosity of the optical counterpart of VCC~2062 a stellar mass of $0.2 \x 10^{8}~\Mo$.  A more realistic value may in principle be obtained from fitting the full Spectral Energy Distribution  with the evolutionary synthesis code PEGASE, as described in Sect.~\ref{sect:stars}. Doing this exercise, we estimate a slightly higher stellar mass, in the range of $0.3-0.7 \x 10^{8}~\Mo$ \citep{Boquien07b}. This estimate is however strongly model dependent and  the star formation history is not very well constrained by the data.
The contribution of the other ingredients -- dust, ionised gas -- is negligible. Depending on the assumed inclination and stellar mass, the total luminous mass accounts for between half and one third of the dynamical mass. Compared to other dwarf galaxies which are known to be dark matter dominated, the \Mdyn/\Mvis\ ratio is low. One should note however that this value has been obtained more or less at the optical radius of the galaxy, while in isolated (massive) galaxies, dark matter becomes dominant at larger radii.

\section{Discussion}
\label{Sec:disc}
In many respects, the properties of VCC~2062 are at odds with those of other dwarf galaxies in the Virgo Cluster.
First of all, it belongs to a large, complex HI structure. With respect to its optical luminosity, its molecular gas content and metallicity are much higher than expected.
Finally, its dynamical to luminous mass is atypical.
We discuss here the various possible origins for VCC~2062, which may account for all these anomalies. We discuss the following hypotheses: VCC~2062 could be a pre-existing dwarf galaxy, a tidally disrupted low surface brightness galaxy or a recycled galaxy made out of material expelled by either ram pressure or tidal forces.

\subsection{A pre-existing dwarf galaxy?}
VCC~2062 has the optical luminosity (\Mb=--13) and even total mass (\Mdyn=$3-4  \x 10^{8}~\Mo$) of a dwarf galaxy, forming stars at a rather modest rate (SFR=$10^{-3}$ ~\uSFR). It could then be one of the numerous ``old'', pre-existing, dwarf irregular galaxies so far discovered in the Virgo Cluster. Would that be the case, it would be very unlikely that all of the HI in which it apparently is embedded would originally have come from the dwarf: it would have had an \MHI/\Lb\ ratio greater than 30, more than 50 times the average value of the Virgo dIrrs \citep{Conselice03}. Besides, the tidal forces exerted during the encounter between a dwarf with the mass of VCC~2062 and a galaxy like NGC~4694 are not strong enough to expel from the latter large quantities of gas. Therefore such a scenario implies that VCC~2062 is by chance currently embedded in an HI tail that had been formed before by an unrelated event.

The strongest arguments put forward to firmly exclude the hypothesis that VCC~2062 is a pre-existing dwarf are its oxygen abundance --- one dex too high with respect to dwarfs of the same luminosity and mass (see Fig.~\ref{fig:met-virgo}) ---, and the strong intensity of its CO lines (see Sect.~\ref{Sect:gas}).

\subsection{A tidally disrupted low surface brightness spiral galaxy?}
If the metallicity and CO content of VCC~2062 are inconsistent with it being a pre-existing dwarf, could it still be the remnant of a pre-existing more massive, metal and gas rich object that would have suffered from a severe tidal disruption? Such a scenario sounds attractive. The collision between NGC~4694 and the progenitor of VCC~2062 may have formed the large HI tail that now links the two galaxies. The low optical luminosity of the remnant could be accounted for if, before the interaction, it was a late-type, low-surface brightness spiral. Dispersed, its old stars would have become invisible on our optical images. This would hence not contradict the results of our SED fitting exercise: the lack of an old stellar population in VCC~2062 (see Sect.~\ref{sect:stars}).

A convincing argument against this scenario is however the small dynamical mass inferred from the rotation curve of the HI still associated with the optical object: \Mdyn=$3-4  \x 10^{8}~\Mo$ is much too low for a spiral, even a small one. Contrary to the stellar component initially located in a disk, there is no easy way to get rid of the dark matter halo in the initial phase of a tidal interaction, and thus the dynamical mass should not dramatically decrease during the collision. In fact, tidal forces perturb the global gas motions and can cause local velocity increases, leading to an over-estimate of the dynamical mass. They also increase the velocity dispersion of the individual clouds \citep{Mihos96,Kronberger06} and the turbulence of the ISM \citep{Elmegreen93}. Last but not least, the rather small widths of the individual HI and CO lines towards VCC~2062, of about 20 \kms\ for the HI component (FWHM) and 15 \kms (FWHM) for the CO component, rather indicate that the local gas is not very perturbed or at least has had time to settle down.

\subsection{A recycled dwarf galaxy?}
If VCC~2062 is not the remnant of a larger galaxy, where does the pre-enriched material it is composed of come from?
A natural candidate is the nearby massive galaxy NGC~4694 to which it is linked by a gaseous bridge. Actually there is no other gas provider in the neighbourhood, as shown in Fig.~\ref{fig:env}. In clusters of galaxies, several mechanisms may in principle contribute to remove gas from a galaxy: ram pressure, galactic harassment and tidal interactions. Once in the intracluster medium, this gas may be recycled to form ``intergalactic stars'', and even, if abundant enough, a new generation of dwarf galaxies.

\subsubsection{Made out of ram pressure stripped gas?}
The interstellar gas of galaxies moving through the intracluster medium feels the pressure of the latter. Such ram pressure strips the galaxies from their gas reservoirs and causes it to be lost to the Intra-Cluster Medium (ICM) in directions opposite to the cluster core. Examples of ram-pressure induced streams of atomic hydrogen are numerous in the Virgo Cluster; the morphology of some of these show some similarities with the long one-sided HI tail in the NGC~4694/VCC~2062 system \citep[e.g.][and references therein]{Vollmer07,Chung07}.
One potential problem with the claim that ram pressure is responsible for the formation of the HI bridge is the location of NGC~4694 in the outermost regions of Virgo, at more than 1 Mpc from the cluster core. The ICM density is locally too low to efficiently remove gas from spiral disks. The confirmed ram-pressure stripped galaxies known in Virgo usually lie in regions where the ICM is denser with some interesting exceptions: \object{NGC~4654} the ISM of which may have suffered weak ram pressure, combined with stronger tidal effects \citep{Soida06} and \object{NGC~4522}, despite lying at more than 1 Mpc from the cluster core, has recently been subject to strong ram pressure, according to radio continuum polarimetry and a model by \cite{Vollmer06}. The proposed scenarios to account for its stripping include a high velocity of the galaxy with respect to the cluster or internal motions within the ICM itself (due to an infalling substructure) which could enhance the ram pressure by a factor of three \citep{Kenney04}. There is however no indication that a similar phenomenon has occurred in the case of NGC~4694.

The main argument against the ram pressure hypothesis is the internal structure of the HI tail which differs from the streams made of ram-pressure stripped gas: they have usually a much lower column density, just reaching the threshold to collapse and form stars \citep{Vollmer03,Kemp05}, but not high enough to form a gravitationally bound object as massive as a dwarf galaxy.

\subsubsection{A tidal origin?}
The high-speed collisions responsible for the so-called galactic harassment process in clusters may produce single-tailed gaseous filaments as shown by \cite{Bekki05} and more recently by \cite{Duc07a}. However a tail as massive and structured as the one emanating from NGC~4694 is best explained by a tidal interaction involving a slow encounter resulting in a merger. VCC~2062 would then be a Tidal Dwarf Galaxy.

Obviously a tidal collision involves two partners, and as previously mentioned, NGC~4694 is the only massive galaxy in the vicinity. The closest one, \object{NGC~4733}, is a low luminosity elliptical galaxy lying at a projected distance of 42' (200 kpc). Thus the tidal hypothesis requires NGC~4694 be an old merger. Indeed, the galaxy exhibits many signs of disturbances that could be explained by a past collision: irregular isophotes, a lopsidedness towards the direction of the HI tail, prominent dust lanes (see Fig.~\ref{fig:env}), an exceptionally small rotation velocity \citep{Rubin99} and recent star formation near the nucleus, as traced by \Ha, far-ultraviolet emission (see Fig.~\ref{fig:system}) and mid-infrared emission \citep{Boselli03}.
Since a stellar disk is still visible, it is unlikely that it is the result of a major merger between equal mass galaxies. 
According to the numerical simulations of \cite{Bournaud06}, collisions between galaxies with mass ratios of 1:5 could at the same time cause relatively strong morphological disturbances in the primary galaxy and form in the tidal tail of the accreted companion a massive tidal object. There is an hint that the building material of VCC~2062 rather came from the accreted companion of NGC~4694:  were  NGC~4694 itself be the parent galaxy, the recycled dwarf should -- statistically -- lie, in projection, closer to the major axis of its progenitor (and thus in 3D near the disk plane), and not perpendicular to it as observed. For any given system, TDGs may end up at a location well above the plane but this situation is very rare \citep{Bournaud06}.

The fact that NGC~4694 exhibits a single nucleus puts some constraints on the age of the collision. A complete merger takes time, typically 0.5 Gyr. Such an age would actually be consistent with the absence of a well defined stellar tail associated with the HI debris. Any tail will have fallen back already or have dispersed. Alternatively, the accreted companion was particularly gas rich and HI dominated in its external regions which are the most affected ones during a tidal interaction.

In many respects, the tidal hypothesis is an attractive one for the origin of VCC~2062. First, general numerical simulations of collisions indicate that it is a viable scenario: TDGs may form in unequal mass mergers out of material pulled out from the accreted companion. Moreover it best explains many observational facts: in particular the morphology of NGC~4694, and the high metallicity and molecular gas content of VCC~2062 with respect to its optical luminosity. The predicted age of the merger (several hundreds of Myr) is actually consistent with the mean age estimated for the background stellar population of the dwarf (see Sect.~\ref{sect:stars}). The ratio between its dynamical and luminous masses, estimated to be of order 2--3, is much lower than for field dwarf galaxies. In fact, TDGs are predicted to be essentially free of dark matter. They are composed of material mainly coming from the rotating disks of their parent galaxies and contain only a tiny fraction of dark matter initially located in their pressure supported halos. The finding that there may be a dark component present in VCC\,2062 is not necessarily a problem. In fact, it fits with the recent discovery by \cite{Bournaud07} of missing mass in the   collisional debris of NGC 5291. Their observations favour the presence of a dark component in spiral disks, possibly in the form of cold molecular hydrogen, that would thus end up in tidal tails, would a collision occur. The missing mass measured in VCC 2062 from the HI data is actually of the same order as in the recycled dwarfs around NGC 5291.

The precise shape of the HI tidal bridge in the NGC\,4694/VCC\,2062 system remains a puzzle. A numerical model of the collision would probably have a hard time matching its many features\footnote{Carrying out such a model would be particularly difficult given the lack of knowledge for many key parameters of the collision. This likely old merger has lost the memory of the initial conditions of the encounter.}. Whereas previous numerical simulations succeeded in reproducing the kinematically decoupled part of tidal tails --- at the origin of TDGs \citep{Duc04b}, and observed towards VCC~2062  --- they rarely form tails as thick as the one associated with NGC~4694. West of the dwarf its extent in the N--S direction reaches more than 10 kpc while the length of the tail is only 30 kpc. Such a low aspect ratio is unusual for a tidal tail but, in principle, could be explained by projection effects. Two branches are actually visible just West of NGC~4694. Unfortunately, the very regular velocity field there -- the two features have similar velocities (van Gorkom et al., in prep.) -- is not really consistent with that predicted by projection effects \citep{Bournaud04}. Clearly additional processes may have occurred to create the puzzling morphology of the tail.
While ram-pressure is unlikely to be at the origin of the tail, as argued previously, it may still have played a role in shaping it.

\section{Conclusions}
We have presented a multi-wavelength analysis of the peculiar elongated HI structure located in the ICM close to NGC~4694, a disturbed early type galaxy located in the outskirts of the Virgo Cluster. The gaseous tail exhibits two main condensations. One of them, with an HI mass of $7.0 \x 10^{7}~\Mo$, appears kinematically detached from the rest of the stream. A well-defined velocity gradient  suggests that the object is gravitationally bound and rotating. A strong CO signal was detected with the IRAM 30m antenna towards the HI peak. We mapped this molecular component which, using a standard galactic conversion factor of $\NHH/\ICO=2\x10^{20}$ \cmm, amounts to about 30\% of the mass of atomic hydrogen. The optical counterpart -- a low surface brightness blue object -- corresponds to the previously identified and catalogued faint dwarf galaxy VCC~2062. An \Ha\ map exhibits three compact star formation knots. We measured the optical spectra of their HII regions and determined from the emission lines an oxygen abundance close to solar.

The properties of VCC~2062 significantly differ from those of other dwarf irregulars studied in the Virgo Cluster.
Its high metallicity and the strong CO signal indicate that it is made of material that has been pre-enriched in a larger galaxy. Its dynamical and structural properties exclude that it is simply the remnant of a tidally destroyed spiral. A much more likely hypothesis is that it is a recycled object made after the gravitational collapse of gas clouds expelled in the intra-cluster medium by a tidal process. The gas provider was most probably the nearby galaxy NGC~4694 or rather a companion which has now merged with NGC~4694. Indeed several hints exist that NGC~4694 is an old merger which accreted several hundred Myr ago a gas-rich companion. According to this hypothesis, VCC~2062 would have been formed out of the clouds lost by the latter galaxy during the tidal interaction. Numerical simulations indicate that such a scenario is plausible and would account for many of the observed properties of the dwarf. An ad hoc model of this old collision will be difficult to carry out since several other processes such as ram pressure might have also played a role.

The discovery of a Tidal Dwarf Galaxy in the Virgo Cluster is interesting on several accounts. First of all, it would prove that such objects may be formed in the cluster environment which is a priori not very favourable for the formation of massive and structured  tidal tails 
\citep{Mihos04} and therefore of Tidal Dwarf Galaxies. Indeed the cluster potential well contributes to disperse the tidal debris which becomes too loose to form a bound object.
Besides, if VCC~2062 is indeed a TDG, it should be a rather  ``old'' one since its parent galaxies have already merged. Numerical simulations predict that a fraction of tidal objects can survive a high infant mortality and those that are not destroyed during the few first hundred Myr after the collision may orbit around their parents for several Gyr, like classical satellites. However, such survivors are difficult to identify; most of the TDG candidates so far observed are still in their infancy. VCC~2062, which was probably formed more than several hundreds of Myr ago, qualifies for being such an old TDG, proving that these objects do exist.
Finally, VCC~2062 would also be one of the nearest TDG so far identified. Its proximity makes it an ideal laboratory to carry out studies requiring a high spatial resolution: among them the search for dark matter in tidal debris which tells about the DM content of spiral disks.

 \begin{acknowledgements}
 This research has made use of the GOLD Mine \citep{Gavazzi03}, SDSS, WIKISKY.ORG and NED Databases.
 Funding for the SDSS and SDSS-II has been provided by the Alfred P. Sloan Foundation, the Participating Institutions, the National Science Foundation, the U.S. Department of Energy, the National Aeronautics and Space Administration, the Japanese Monbukagakusho, the Max Planck Society, and the Higher Education Funding Council for England.
 The NASA/IPAC Extragalactic Database (NED) is operated by the Jet Propulsion Laboratory, California Institute of Technology, under contract with the National Aeronautics and Space Administration. We are particularly grateful to the VIVA team, especially to Aeree Chung, who sent us a preliminary HI datacube before the VLA data became public. We thank Jacqueline van Gorkom, Jeffrey Kenney and Bernd Vollmer for their feedback on the paper.
 This work has benefited much from the input of Fr\'ed\'eric Bournaud on the dark matter content of VCC~2062 and Giovanna Temporin who obtained and reduced the Asiago/Copernicus image of the system.  We thank our referee, F. La Barbera, for his careful reading of the manuscript and suggestions for clarification.  UL acknowledges support by the Spanish Ministry of Education and Science via projects AYA2004-08251-C02-00, ESP2003-00915 and by the Junta de Andaluc\'{\i}a.

\end{acknowledgements}
 
 \bibliographystyle{aa}
\bibliography{../../Biblio/all}

\end{document}